\documentclass[
reprint,
amsmath,amssymb,
aps,
prl,
]{revtex4-2}

\usepackage{graphicx}
\usepackage{dcolumn}
\usepackage{bm}
\usepackage{hyperref}
\usepackage{graphicx} 
\usepackage{xcolor}
\usepackage{makecell}
\usepackage{amsmath}
\usepackage{orcidlink}
\usepackage{booktabs}

\mathchardef\mhyphen="2D

\begin{document}

\title{AI-enhanced tuning of quantum dot Hamiltonians toward Majorana modes}

\author{Mateusz Krawczyk
\orcidlink{https://orcid.org/0000-0003-0466-814X}}
\affiliation{Institute of Theoretical Physics, 
Wroc{\l}aw University of Science and Technology,
Wybrze\.{z}e Wyspia\'{n}skiego 27,
50-370 Wroc{\l}aw, Poland}

\author{Jaros\l{}aw Paw\l{}owski
\orcidlink{0000-0003-3638-3966}}
\email{jaroslaw.pawlowski@pwr.edu.pl}
\affiliation{Institute of Theoretical Physics, 
Wroc{\l}aw University of Science and Technology,
Wybrze\.{z}e Wyspia\'{n}skiego 27,
50-370 Wroc{\l}aw, Poland}

\begin{abstract}
We propose a neural network-based model capable of learning the broad landscape of working regimes in quantum dot simulators, and using this knowledge to autotune these devices -- based on transport measurements -- toward obtaining Majorana modes in the structure. The model is trained in an unsupervised manner on synthetic data in the form of conductance maps, using a physics-informed loss that incorporates key properties of Majorana zero modes. We show that, with appropriate training, a deep vision-transformer network can efficiently memorize relation between Hamiltonian parameters and structures on conductance maps and use it to propose parameters update for a quantum dot chain that drive the system toward topological phase. Starting from a broad range of initial detunings in parameter space, a single update step is sufficient to generate nontrivial zero modes. Moreover, by enabling an iterative tuning procedure -- where the system acquires updated conductance maps at each step -- we demonstrate that the method can address a much larger region of the parameter space.
\end{abstract}

\maketitle

\section{Introduction}

Majorana zero modes (MZMs) are emergent quasiparticles predicted to obey non-Abelian statistics, enabling topologically protected quantum information processing~\cite{Alicea2012,Beenakker2013,Sau2010,Oreg2010}. Their realization in hybrid superconductor-semiconductor systems requires precise control over local parameters such as chemical potentials, inter-dot couplings, and spin-orbit interactions~\cite{Leijnse2012,Lutchyn2018}. Small deviations from the so-called \textit{sweet-spot} conditions of the Kitaev chain (KC)~\cite{Leumer2021} can destroy the topological gap and delocalize MZMs~\cite{Kitaev2001}. Moreover, it is not trivial to distinguish between MZMs and different nontopological zero-bias peaks ~\cite{Aguado2021,Sasaki2000,Pikulin2012,Kells2012}. Consequently, achieving robust Majorana states experimentally remains a major challenge, particularly in the presence of fabrication disorder and parameter noise.

To address this challenge, Fulga \textit{et al.}~\cite{Fulga2013} proposed an adaptive tuning protocol for superconductor-proximitized chain of quantum dots (QDs) that simulates a KC generalized on non-uniform parameters distribution across dots, trying to demonstrate how gate-controlled voltages and superconducting phases can be adjusted to reach a topologically nontrivial regime. In this paper we revisit this problem trying to automate it using deep neural estimators.
More recently, further works indicate that through appropriate tuning, it is possible to realize a KC with emerged MZMs using QDs coupled through elastic co-tunneling and crossed Andreev reflection~\cite{Leijnse2012a,Leijnse2012, Sau2012, Dvir2023, Bordin2025}. All these proposals utilize coupling with more readily available $s$-wave superconductors; however, there are also proposals employing $p$-wave superconductors~\cite{Mohseni2023,Ezawa2024}.

At the same time, a different strategy for machine learning (ML)-assisted autotuning of QD-based quantum simulators~\cite{Borsoi2024,Mills2019,Shandilya2025} using transport measurements is gaining significant interest~\cite{Darulov2020,Lunczer2019,Zwolak2023,roux2025,losert2025}. The use of transport measurements in the form of \textit{conductance maps}~\cite{Donarini2024,Shandilya2025} to derive insights about the system Hamiltonian~\cite{Blonder1982,grepkowa2023, Wang2017, Gebhart2023,Thamm2024,Taylor2024disorderlearning,Taylor2025analysis} appears to be a promising path towards the automatic tuning of its parameters. In particular, inverting the measured conductance matrices to determine electrostatic potential disorder, using evolutionary optimization~\cite{Thamm2024}, supervised neural networks (NNs)~\cite{Taylor2024disorderlearning,vandriel2024,Taylor2025analysis}, or both~\cite{Taylor2025disordermitigation}, is a natural first step that can allow precise tuning of Hamiltonian parameters towards MZMs. Although these methods already allow for successful mitigation of disorder in nanowires \cite{Taylor2025disordermitigation,Thamm2024}, they rely on indirect, non-differentiable cost functions and heuristic evolutionary searches, rather than learning the broader behavior of the underlying physical system.

\begin{figure}[!b]
    \includegraphics[width=.98\linewidth]{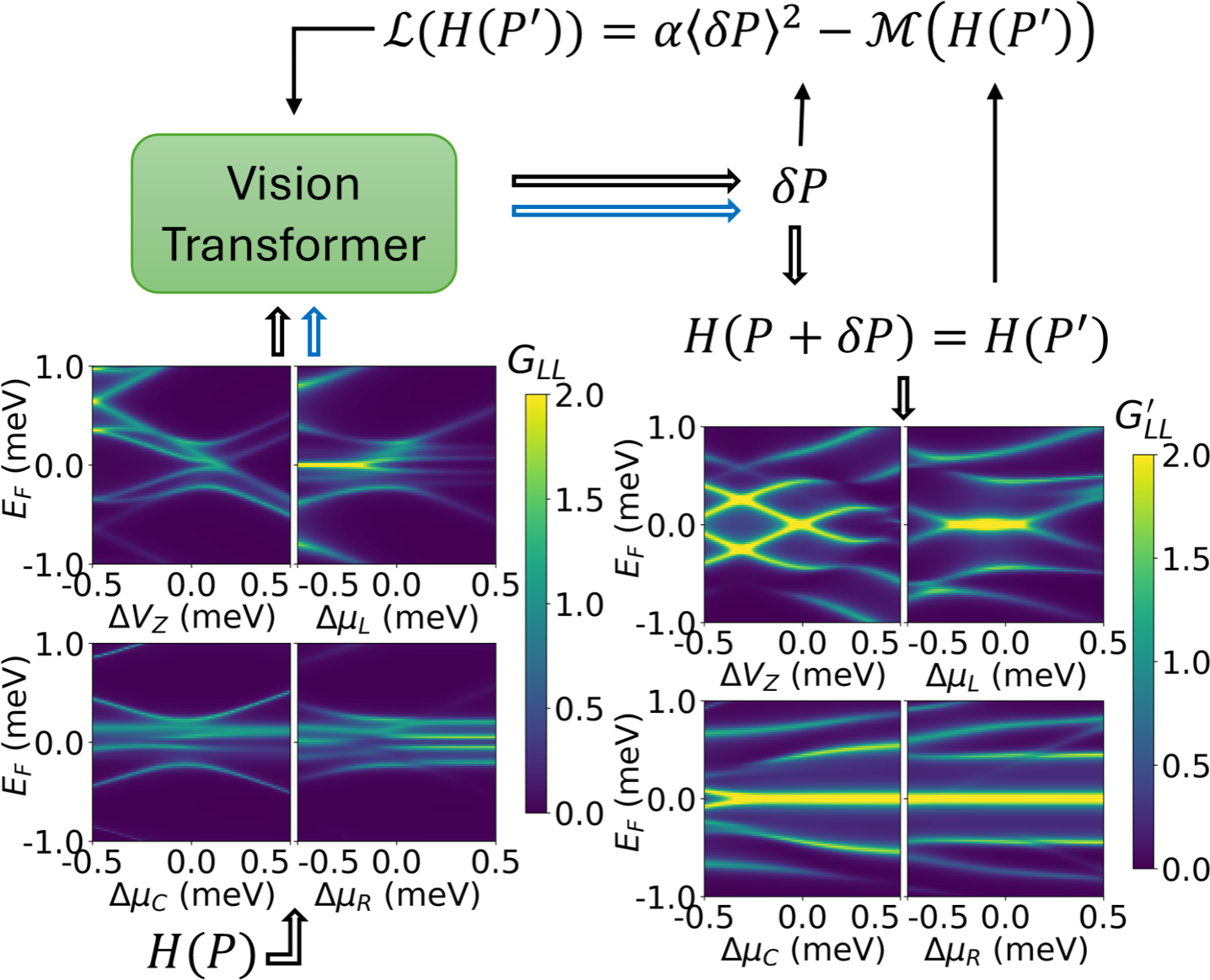}
    \caption{Scheme of the QDs-based KC-simulator autotuning system (PINNAT). Black (blue) arrows show the system training (inference) path. Vision transformer model is fed with conductance maps to predict parameter corrections, that should decrease $\mathcal{M}$-based loss function, simultaneously increasing the probability of MZMs emergence.}
    \label{fig:scheme}
\end{figure}
In contrast, here we introduce an unsupervised \textit{physics-informed}~\cite{Raissi2019,kovachki2023neural} NN-based auto-tuning (PINNAT) model with Majorana physics embedded in the loss function, and show that by proper training the model can memorize various structures on conductance maps which enables to build an efficient quantum simulator autotuning system. Our AI-enhanced adaptive tuning framework directly extends QDs simulator by leveraging \textit{vision transformer} (ViT)~\cite{dosovitskiy2021} NN architecture.
The scheme of the proposed approach is presented in Fig.~\ref{fig:scheme} with information flow during the model training (black arrows) and evaluation (\textit{inference}, blue arrows).

\section{Model}

Following a lattice model from Ref.~\cite{Fulga2013}, the effective superconductor-proximitized Rashba-Zeeman Hamiltonian for a chain of spinful single-level QDs can be expressed as follows
\begin{align}
H = {} &
\sum_{n,s,s'}
\Bigl[
(-\mu_n \sigma_0 + V_\mathrm{Z} \sigma_z)_{ss'} \, c_{n,s}^\dagger c_{n,s'}
\Bigr. \nonumber \\[4pt]
& \left.
+ \tfrac{1}{2} 
\left(
\left(\Delta_{n}e^{i\phi_n} i\sigma_y \right)_{ss'} 
c_{n,s}^\dagger c_{n,s'}^\dagger 
+ \mathrm{h.c.} 
\right)
\right. \nonumber \\[4pt]
& +\Bigl.\left(
t_n (e^{i\boldsymbol{\lambda}_n \cdot \boldsymbol{\sigma}})_{ss'} 
c_{n,s}^\dagger c_{n+1,s'} 
+ \mathrm{h.c.}
\right)
\Bigr],
\label{hamiltonian}
\end{align}
with the on-site potentials \(\mu_n\), inter-dot hopping amplitudes \(t_n\) modulated by (Rashba) spin-orbit vector \(\boldsymbol{\lambda_n}=\lambda_n(\sin\rho_n\cos\xi_n, \sin\rho_n \sin\xi_n, \cos\rho_n)\), Zeeman energy \(V_\mathrm{Z}\), proximity-induced superconducting ($s$-wave) pairing \(\Delta_ne^{i\phi_n}\), and $n$ describing dot number while $s$ being spin degree of freedom.
The values of those parameters are then carefully adjusted to have MZMs emerge in the system. One set of tuned (yet uniform) parameters (which we call \textit{reference parameters}, $P_0$) for $N=3$ QDs chain includes: $\mu=0.6\,\mathrm{meV}$, $t=0.25\,\mathrm{meV}$, $\lambda=0.27\,\pi$, $\rho=\xi=\frac{\pi}{2}$, $V_\mathrm{Z}=0.5\,\mathrm{meV}$, and $\Delta=0.25\,\mathrm{meV}$.
The Rashba length $\lambda = 0.27\,\pi$ was tuned so that at $\mu = 0.6$~meV two energy levels touch at zero energy (c.f. Fig.~\ref{fig:sample_lambda}(a)).
Also note that some parameters, i.e., $\{\mu_n$, $t_n$, $\lambda_n\}$ (7 in total) are easier to tune -- electrically (via local gating) than the others: the global Zeeman field $V_\mathrm{Z}$ and the proximity-induced superconducting gap $\{\Delta_n\}$.

To train the ViT model in physics-informed manner, we introduce a differentiable quasi-metric \(\mathcal{M}\), quantifying how close a given Hamiltonian is to MZM regime. The proposed phenomenology combines several physical indicators: edge-state localization, zero-energy spectral weight, and parity (electron-hole) symmetry. For a detailed $\mathcal{M}$ definition, see the Methods section. We also note that we tested the popular Majorana \textit{polarization}~\cite{Sticlet2012,Setiawan2015,Maska2017,Szumniak2017} measure, directly related to the parity operator, but it yielded unsatisfactory training results.
This is because it fails to discriminate topologically trivial zero modes that localize in the central QD or anti-localize in the left and right QDs (see the Supplementary Information of Ref.~\cite{Bordin2025} for examples of such states). We define our measure in a way that also discriminates these states.
Another approach to detecting the presence of MZMs is the recently proposed protocol based on quantum Fisher information~\cite{plodzien2026}; however, its effectiveness in our case remains to be verified.

We start with random set of parameters $P$ and collect conductance maps $G(H(P))$ -- details on $G$ calculation can be found in the Methods section.
PINNAT fed by $G$ maps predicts corrective updates $\delta P$ to a subset of Hamiltonian parameters. We assume that, depending on the experimental setup, different subsets of parameters may be available for tuning. Two versions of the PINNAT model to mimic different experimental arrangements was trained. The first one is trained to predict corrections to electrically controlled local parameters \(\{\mu_n, t_n, \lambda_n\}\) -- each of which can be adjusted independently, while leaving \(\{\Delta_n, V_\mathrm{Z}\}\) as background variables. The second one is allowed to adjust \(\{\mu_n, V_\mathrm{Z}\}\) pair, where $\mu_n$ can be corrected locally, while $V_\mathrm{Z}$ is adjusted globally. We also have to keep in mind that PINNAT model (as well as the experimentalist) \textit{does not know} which parameters were detuned.

Maximizing \(\mathcal{M}\) drives the NN to adjust $P$ such that edge states become increasingly localized and the strong mid-gap energy signal emerges, which is visualized in Fig.~\ref{fig:scheme} by output conductance maps (this example is further analyzed in Fig.~\ref{fig:sample_lambda}(d)). Specifically, PINNAT is trained to minimize the loss:
\begin{equation}
\mathcal{L}(H(P')) = \alpha \left\langle\delta P\right\rangle^2 - \mathcal{M}(H(P')),
\label{eq:loss}
\end{equation}
for the Hamiltonian with tuned parameters \(H(P')=H(P\,+\,\delta P)\). Additionally, to force the PINNAT to predict smallest possible correction \(\delta P\) to the parameters $P$, an extra regularization term $\alpha\langle\delta P\rangle^2$ is included, with a factor \(\alpha=0.1\).
Training details, including parameters sampling used to generate synthetic training set of conductance maps, are described in the Methods section.

\section{Results}
\begin{figure}[!htb]
    \includegraphics[width=.99\linewidth]{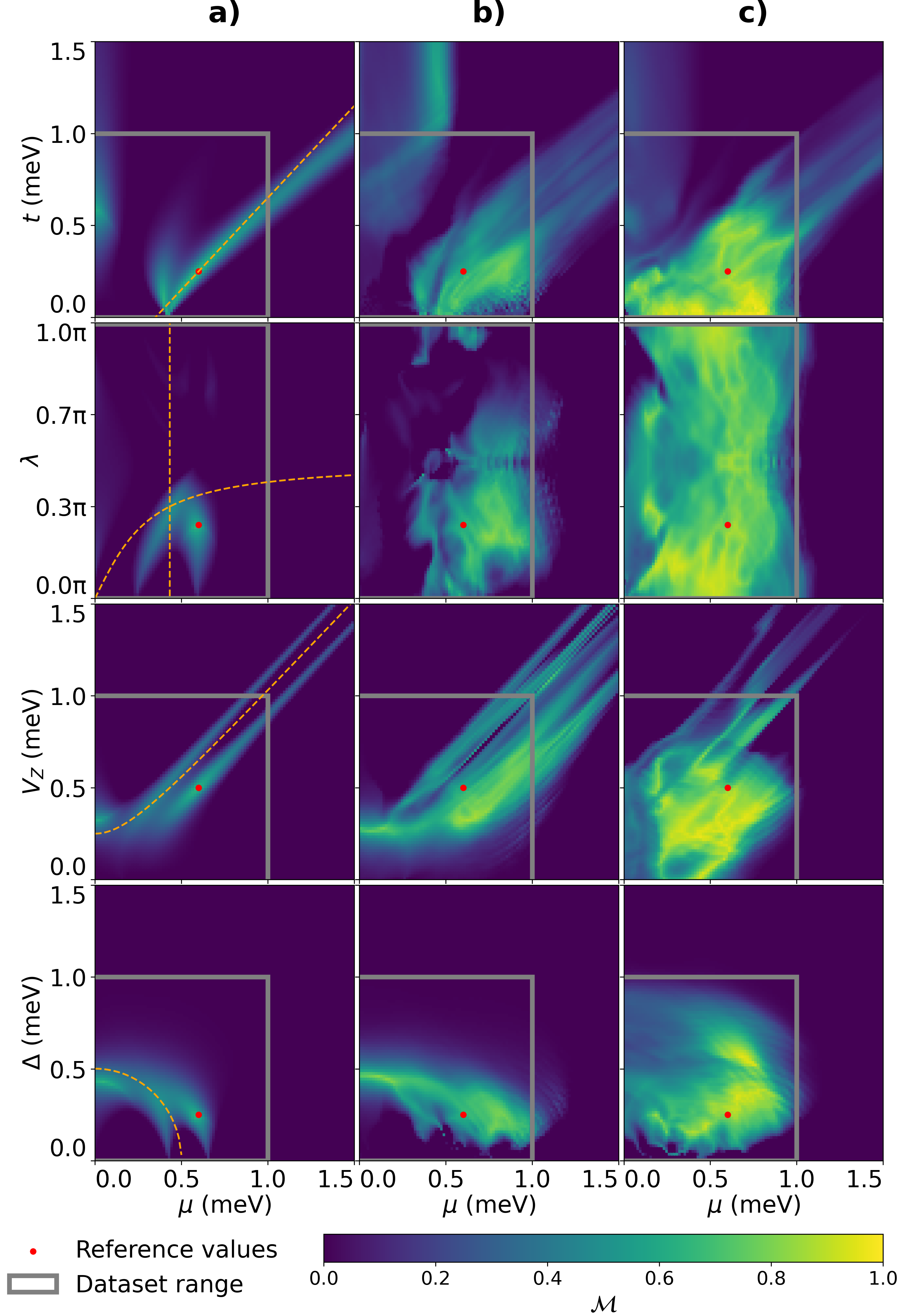}
    \caption{$\mathcal{M}$ metric map for uniform (across the dots) change of different parameters -- resulting in shifted $P$ -- for (b) model adjusting \(\{\mu_n, t_n, \lambda_n\}\), and (c) adjusting \(\{\mu_n, V_\mathrm{Z}\}\). The ranges of the parameter sampling that were used in the model training set are marked with a gray rectangle. The reference parameter values $P_0$ are marked with a red dot. In (a), we present values of $\mathcal{M}(H(P))$ before parameter tuning, while in (b) and (c) $\mathcal{M}(H(P+\delta P))$ -- after tuning.}
    \label{fig:m_map_total}
\end{figure}

To present the PINNAT performance in parameter tuning, we plot the values of $\mathcal{M}$, disturbing selected three-QDs Hamiltonian parameters. Specifically, we shift (detune) the value of the given parameters (the rest is kept default, $P_0$) -- obtaining $P$ -- and verify what corrections $\delta P$ are predicted by the NN models. Then the $\mathcal{M}(H(P+\delta P))$ value is plotted.
In Fig.~\ref{fig:m_map_total}(b) and (c) we show how $\mathcal{M}$ changes with uniform (i.e., the same for each QD) parameter shift for models tuning \(\{\mu_n, t_n, \lambda_n\}\) and \(\{\mu_n, V_\mathrm{Z}\}\), respectively.
The values for $\mathcal{M}(H(P))$, i.e. before correction, are presented as reference in column (a).  
For example, in $t$ vs $\mu$ plot we shift parameters to $t_1=t_2=t$ and $\mu_\mathrm{L}=\mu_\mathrm{C}=\mu_\mathrm{R}=\mu$ obtaining $P$ (column (a)), then predict correction $\delta P$ and present $\mathcal{M}(H(P+\delta P))$ in respective positions in columns (b) and (c).
Additionally, in column Fig.~\ref{fig:m_map_total}(a), orange dashed curves mark the analytical conditions:
(1) $\Delta=\sqrt{V^2_\mathrm{Z}-\mu^2}$
for the existence of two zero modes in decoupled QDs, and (2) $\lambda=\arctan(\frac{\mu}{\Delta})$ corresponding to the KC \emph{sweet-spot} (see the Appendix~A for details).

Results in In Fig.~\ref{fig:m_map_total}(b) and (c) show that the PINNAT models can effectively learn to identify various $H(P)$ regimes from $G$ maps and use it to significantly increase -- in comparison to Fig.~\ref{fig:m_map_total}(a) -- the regions with $\mathcal{M}>0$ by proposing corrections $\delta P$ to the subset of the parameters. Unsurprisingly, the models are most effective in the parameter regime covered in the training data, although some ability to \emph{generalize} on unseen parameter ranges can be noticed, especially for $t$ and $V_\mathrm{Z}$ shift in Fig.~\ref{fig:m_map_total}(b). 
On the other hand, the PINNAT that updates the \(\{\mu_n, t_n, \lambda_n\}\) exhibits difficulties in some regions in predicting appropriate corrections for the values of $t$ and $\lambda$, even though they are allowed to be modified.

Interestingly, the model that adjusts \(\{\mu_n, V_\mathrm{Z}\}\), presented in Fig.~\ref{fig:m_map_total}(c), is slightly more effective in proposing corrections, both in terms of parameter range that can be effectively corrected and the probability of MZMs emergence, as indicated by higher values of $\mathcal{M}$. This behavior is specifically worth highlighting as the model does not modify the $t_n$, $\lambda_n$ and $\Delta$ parameters, while still possessing the ability to effectively amend them by updating $\mu_n$ and $V_\mathrm{Z}$.
These results show that $V_\mathrm{Z}$ 
plays a more important role in controlling the system than the hoppings $t_n$ and SOI $\lambda_n$. This is consistent with observations known from Rashba wires~\cite{Alicea2012,stanescu2024}.
Having a (perpendicular) magnetic field removes the so-called fermion doubling problem by opening a gap (at $k=0$). The size of this gap increases with the Zeeman field, which means that for larger Zeeman splitting there is a wider range of $\mu_n$ for which the nontrivial phase is present. This is what we observe.
However, it should be noted that when the hopping parameters $t$, $\lambda$ are  close to zero in the first and second rows of Fig.~\ref{fig:m_map_total}(c) the model likewise selects symmetric zero modes localized at the edge QDs with a high $\mathcal{M}$; nevertheless in this case (in the limit $t\rightarrow 0$) the system is clearly in a trivial phase~\cite{stanescu2024}.
For $\lambda=0$ there is no possibility of realizing $p$-wave pairing in the KC, and the entire model breaks down.
\begin{figure}[!tb]
    \includegraphics[width=.99\linewidth]{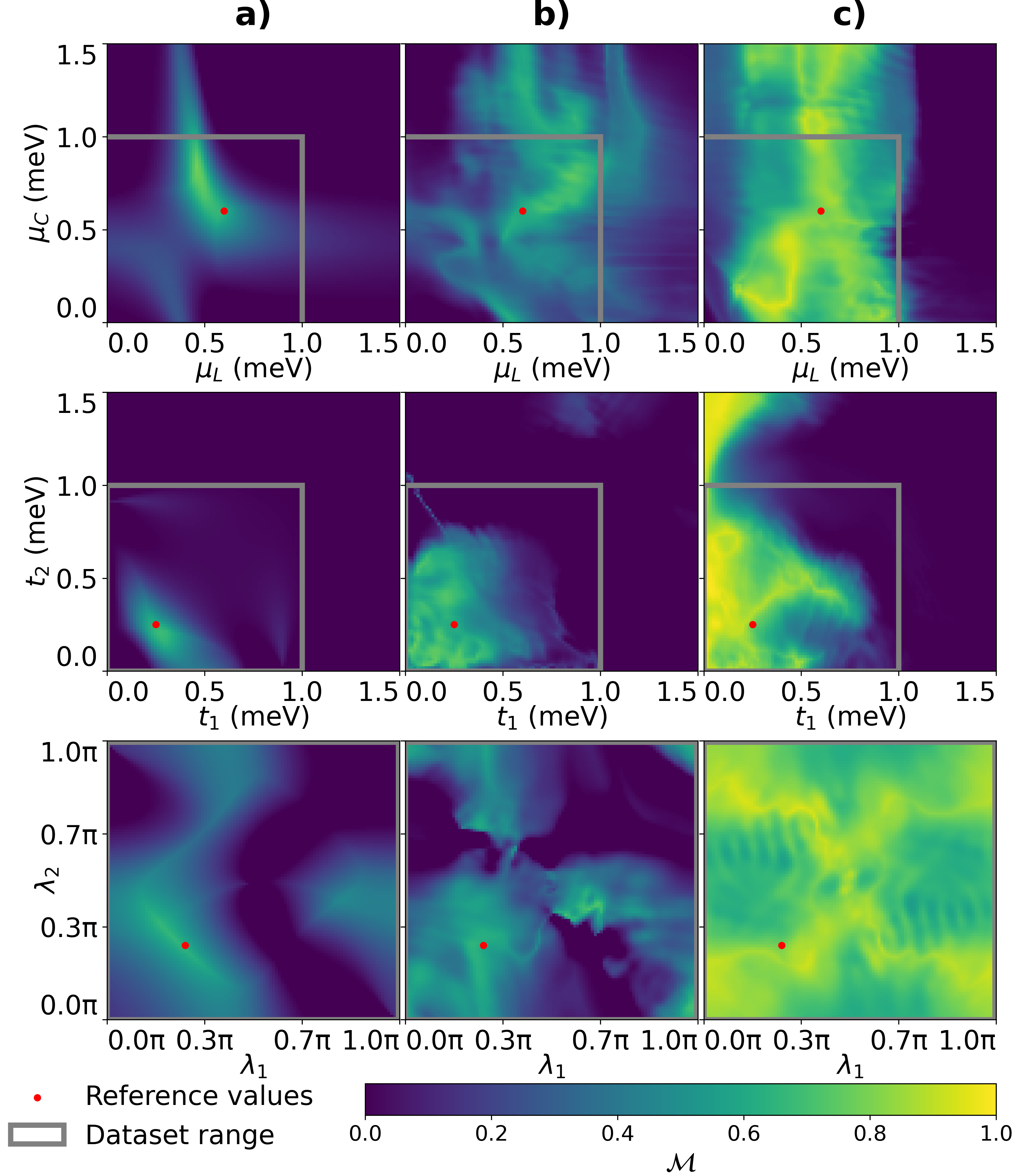}
    \caption{Same as in Fig.~\ref{fig:m_map_total}: $\mathcal{M}$ metric maps (a) before, and (b,c) after parameters tuning, but now for \textit{local} shift of selected pair of parameters.}
    \label{fig:m_map_dots}
\end{figure}
One way to mitigate this behavior is to include regularization terms in the training phase that push these parameters (t and $\lambda$) away from zero. However, to simplify the training procedure, we decided not to include it in this work.

PINNAT behavior is further investigated in Fig.~\ref{fig:m_map_dots} showing how $\mathcal{M}$ changes with (a pair of selected) local parameters shift. In case of a shift in the $\mu_n$ parameter, both models produce comparable results -- although tuning \(\{\mu_n, t_n, \lambda_n\}\) allows one to cover a wider range of parameters, adjusting \(\{\mu_n, V_\mathrm{Z}\}\) allows to reach higher values of $\mathcal{M}$. Simultaneously, sweeping $t_n$ and $
\lambda_n$ produces larger discrepancies in favor of the PINNAT model that corrects the parameters \(\{\mu_n, V_\mathrm{Z}\}\). Nevertheless, both models demonstrate the ability to effectively correct the dot-specific parameters and increase the probability of observing desired MZM modes.

\begin{figure}[!tb]
    \includegraphics[width=.99\linewidth]{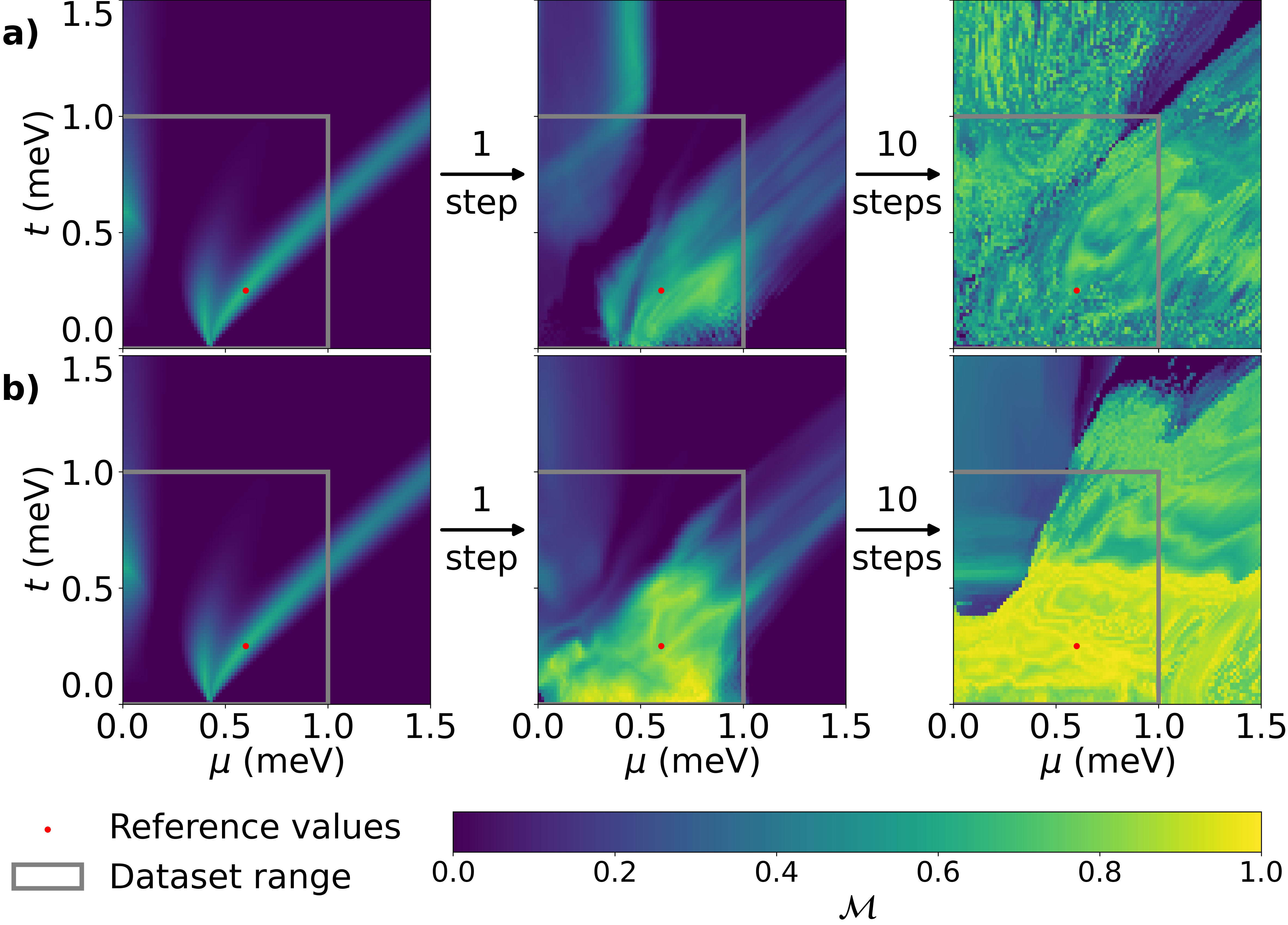}
    \caption{Iterative autotuning procedure for (a) model adjusting \(\{\mu_n, t_n, \lambda_n\}\), and (b) adjusting \(\{\mu_n, V_\mathrm{Z}\}\).  In the first (left) plot, we present $\mathcal{M}$ map before tuning. Then, in the center, parameters are tuned with a single step of NN corrections. Finally, in the last (right) plot, there is a map plotted for parameters tuned within 10 subsequent steps of NN corrections.}
    \label{fig:m_map_autotuning}
\end{figure}

Since the models tend to work in a rather limited area of parameters covered in the training data, we propose an iterative modification to the proposed procedure. After applying the initial corrections proposed by PINNAT, one can measure the conductance maps for the corrected system and pass them to the NN in the subsequent step of autotuning. As a result, in Fig. \ref{fig:m_map_autotuning}, we can notice that within 10 steps of corrections, the $\mathcal{M}$ metric can be significantly increased even in the regions with an initially zero value of $\mathcal{M}$.

\begin{figure}[!tb]
    \includegraphics[width=.99\linewidth]{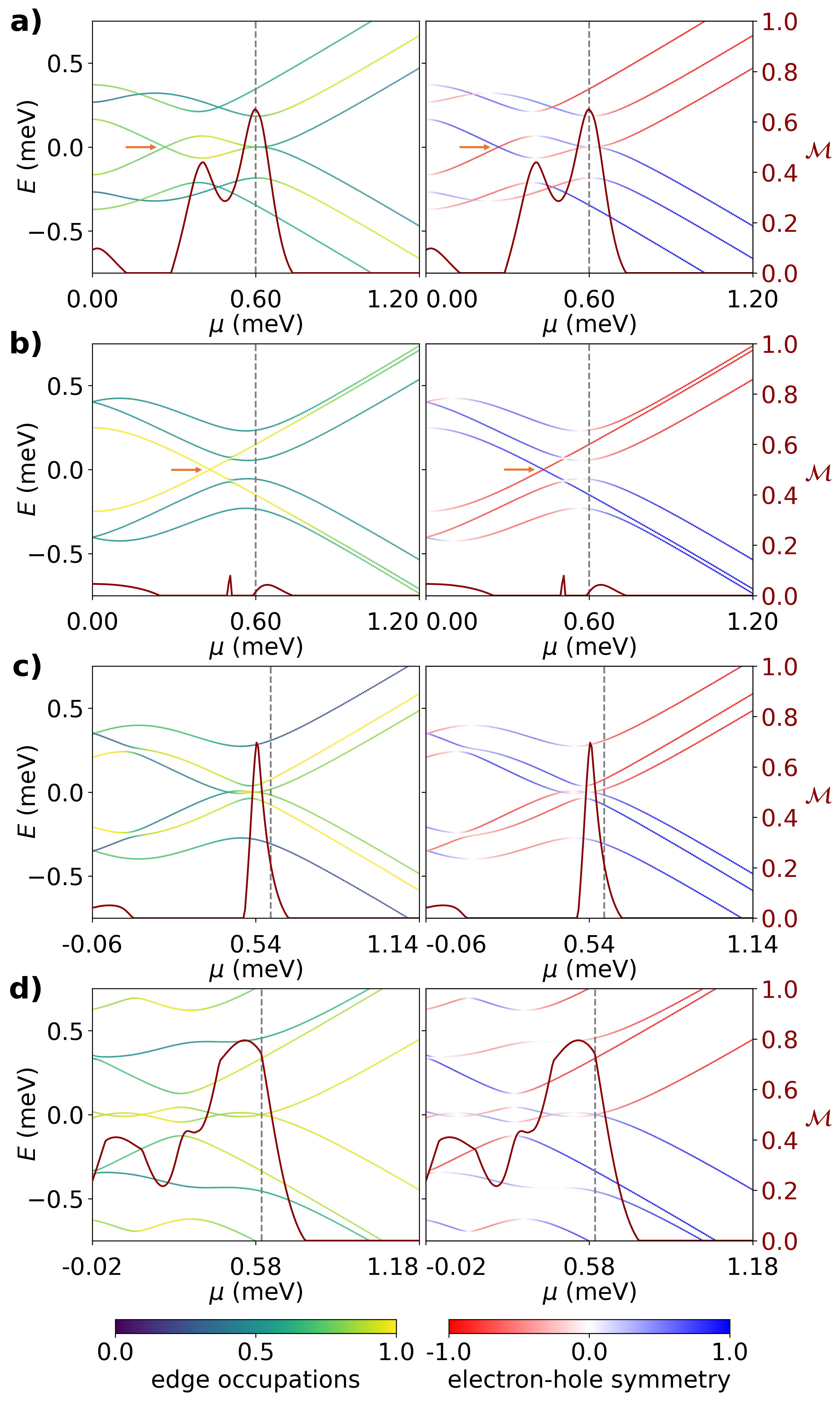}
    \caption{Three-QD chain Hamiltonian $H$ as a function of the offset $\mu$ added to local potentials: $\mu_n\rightarrow\mu_n+\mu-0.6$~meV (dashed vertical line marks the reference $\mu=0.6\,\mathrm{meV}$).
    Additionally, the $\mathcal{M}$ metric is plotted, and eigenvalues are colored (left column) with edge occupations $O_{edge}^{(i)}$, or (right column) electron-hole symmetry $S_{\mathrm{eh}}^{(i)}$ defined in the Methods section.
    Plots are presented for (a) reference parameters; (b) modified $\lambda_1$ and $\lambda_2$; (c) parameters from (b) but with NN-tuned \(\{\mu_n, t_n, \lambda_n\}\); (d) parameters from (b) with NN-tuned \(\{\mu_n, V_\mathrm{Z}\}\). Orange arrow points to the trivial zero modes.}
    \label{fig:sample_lambda}
\end{figure}
\begin{figure}[!tb]
    \includegraphics[width=.9\linewidth]{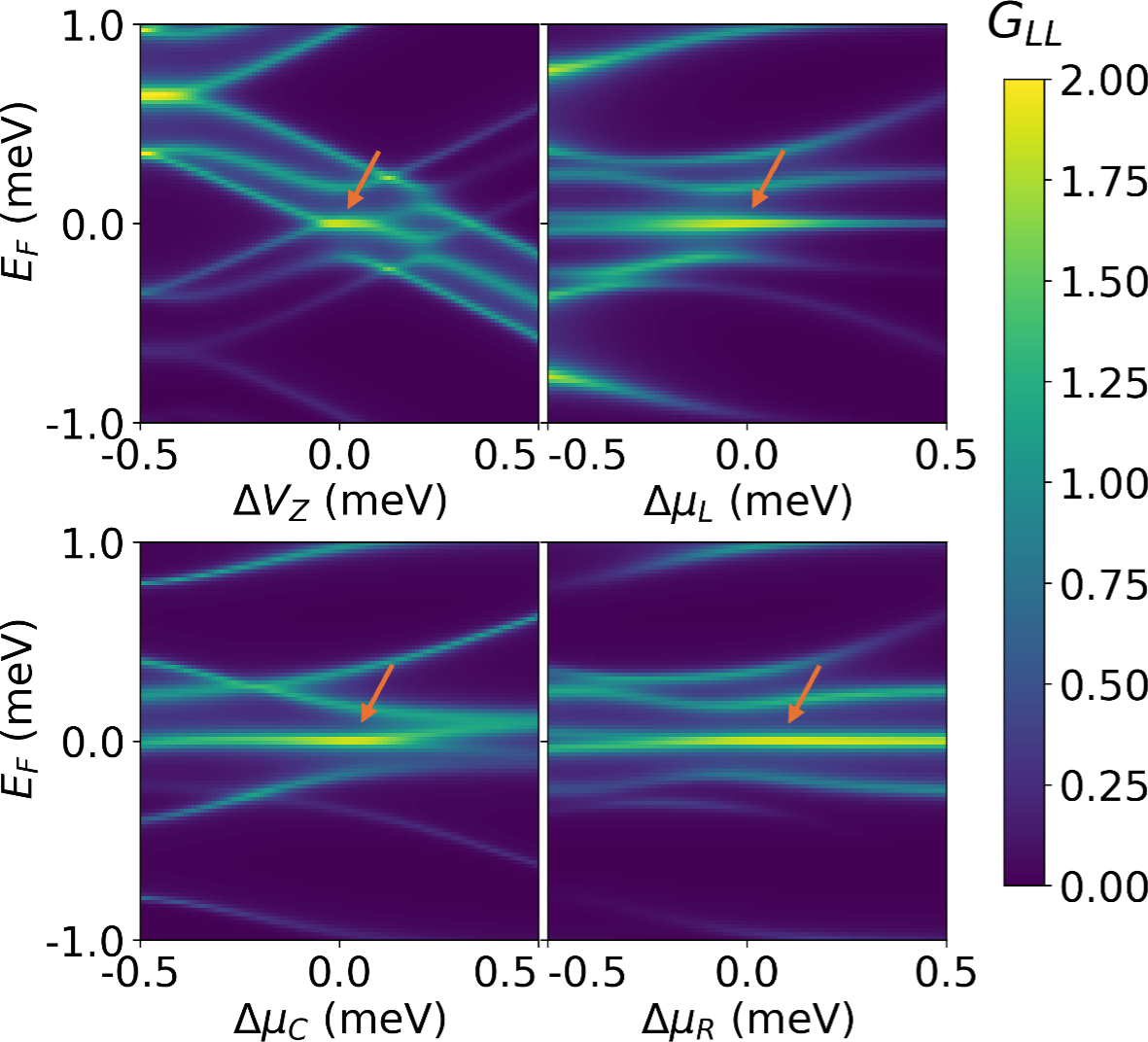}
    \caption{Conductance maps for the reference parameters $P_0$. Zero-bias peaks are indicated by orange arrows.}
    \label{fig:default_conductance}
\end{figure}

Finally, we present a specific example of the autotuning procedure for the $\lambda_n$ parameter. In Fig.~\ref{fig:sample_lambda}, we subsequently plot the eigenvalue spectrum of the three-QD chain Hamiltonian $H$ as a function of the global offset $\mu$ added to local potentials $\mu_n+\mu$.
In Fig.~\ref{fig:sample_lambda}(a) we have exactly reference parameters $H(P_0)$ case, where we can clearly observe MZMs separated (by a
topological gap) from the rest of the spectrum with maintained electron-hole symmetry and (partially) localized on the edges.
Conductance maps for the reference parameters, $P_0$ case: $G(H(P_0))$ are presented in Fig.~\ref{fig:default_conductance}.

Adding a noise to $\lambda_n$ (by setting $\lambda_1=0.62\,\pi$ and $\lambda_2=0.41\,\pi$) -- presented in Fig.~\ref{fig:sample_lambda}(b) -- results in reduced $\mathcal{M}$ value and vanishing MZMs.
Then the noisy parameter can be corrected using either of the two presented NN models. Tuning \(\{\mu_n, t_n, \lambda_n\}\) -- presented in Fig.~\ref{fig:sample_lambda}(c) -- makes MZMs reappear with desired properties, i.e., they are localized on the edges, electron-hole symmetry is preserved, and the edge states are slightly gapped from the rest of the spectrum. On the other hand, tuning \(\{\mu_n, V_\mathrm{Z}\}\) -- Fig.~\ref{fig:sample_lambda}(d) -- not only leads to restoring MZMs at the exact point in the tuned parameter space with the larger topological gap but also increases the probability of observing MZMs in a wider range of $\mu$ values. 
An additional example of analogous autotuning, but for the $t$ parameter is provided in the Appendix~B.

Importantly, Fig. 5 shows that the $\mathcal{M}$ metrics achieve high values only for the topological states. Specifically, trivial zero modes like the Andreev bound state (ABS)~\cite{Liu2017, Prada2020} indicated by the orange arrow in Fig.~\ref{fig:sample_lambda}(a) and Fig.~\ref{fig:sample_lambda}(b) do not increase the $\mathcal{M}$ value. Since the PINNAT model is trained to increase $\mathcal{M}$, only non-trivial zero-energy states emerge after parameter tuning, highlighting the robustness of the proposed approach in distinguishing between trivial and topological states.

\section{Discussion}

These findings align with and extend current efforts in ML-assisted control of Majorana platforms~\cite{Thamm2024,Taylor2025disordermitigation,vandriel2024}. Most notably, the recent work of Taylor and Das Sarma, Ref.~\cite{Taylor2025disordermitigation} introduced a ViT-based framework for mitigating strong disorder in Majorana nanowires. Their study demonstrated that disorder can be compensated -- sometimes restoring topology in systems that originally show no nontrivial regions -- using conductance-based NN inference combined with evolutionary optimizer (CMA-ES). Our approach differs in several important respects. First, instead of relying on a second-stage optimizer such as CMA-ES, our network learns Hamiltonian behaviors manifested by transport measurements to \emph{directly} predict parameter corrections, making the tuning process faster and conceptually closer to a experimental tuning protocol. Second, whereas Ref.~\cite{Taylor2025disordermitigation} focuses primarily on restoring scattering-invariant and LDOS-based indicators in bulk nanowires, our work addresses \emph{QD chains}, a complementary Majorana platform with discrete QDs and gate-defined tunnel couplings between them. Third, our method integrates NN physics-informed via differentiable quasi-metric $\mathcal{M}$ capturing edge localization, electron-hole symmetry, and spectral proximity to zero energy; this contrasts with supervised learning to predict topology-based indicators.

It is instructive to compare the parameter regimes identified by PINNAT with theoretical sweet-spot relations discussed in Appendix~A. The parameter sets obtained in our NN-based autotuning follow the same qualitative trends, namely Zeeman energy exceeding the inter-dot couplings and remaining comparable to $\sqrt{\Delta^2 + \mu_n^2}$, while spin-orbit coupling following the expected sweet-spot relation with onsite potential. These relations are satisfied at the level of order-of-magnitude and averaged parameters, rather than exactly on each site, reflecting the fact that PINNAT optimizes the metric $\mathcal{M}$ instead of enforcing analytical constraints explicitly. A detailed comparison is provided in Appendix~D.

We also emphasize that our results highlight the possibility of \emph{generalization beyond the training regime}. For several parameters, notably $t$ and $V_\mathrm{Z}$, the networks propose meaningful corrections even when initial values fall outside the training set distribution. Specifically, after increasing the number of consecutive steps of corrections, the network is capable of effective autotuning in a broad area of parameters. This indicates that unsupervised PINNAT architecture is capable of learning physically sensible parameter correlations rather than just approximating between samples (Hamiltonians) from the training set. Such generalization is essential if these methods are to be deployed on real devices, where fabrication imperfections and drift may move operating points far from any simulated training set.

Importantly, the proposed PINNAT scheme is fully physics-informed: MZM metric $\mathcal{M}$, together with the Hamiltonian family definition and eigensolver implementation, is explicitly encoded in the structure of the NN loss function. This approach aligns with the recent strong trend toward designing NNs that respect the underlying physics of the problem~\cite{Bronstein2021}, for example in studies of quantum entanglement~\cite{Pawlowski2024,Krawczyk2024}, quantum control~\cite{Norambuena2024}, or in the optimization of quantum tomography protocols~\cite{Krawczyk2025}. Moreover, we argue that similar approaches based on conductance map inputs could be developed to optimize a broad class of physics-informed metrics. In particular, since direct estimation of Hamiltonian parameters from conductance maps appears feasible~\cite{Pawłowski2026}, a wide range of parameter-tuning problems could potentially be addressed using PINNAT-like models.
Tuning the QD system into the single-electron (Coulomb blockade) regime could constitute a standard application in this context.

While our current proof-of-concept study is restricted mainly to a three-QDs system, we show the ability to scale the proposed method to a 7-QDs chain in the Appendix~C. With some optimization adjustments in the training procedure, we claim that it is possible to have the PINNAT model trained to propose significant corrections in the larger systems, maintaining its ability to autotune the Hamiltonian parameters beyond the training regime.

Although our demonstrations focus on QD chains, the PINNAT framework is not restricted to this platform. Since proximitized semiconductor nanowires are routinely modeled using discretized lattice Hamiltonians of the same type~\cite{Lutchyn2018,Leumer2021}, the proposed approach could be extended to nanowire systems---albeit with a limited number of lattice sites, as discussed in Appendix~C---by replacing the underlying Hamiltonian and adapting transport formalism.

The major intrinsic limitation of this work is that our method relies on simulated conductance maps, whereas experimental data may include additional noise sources, nonidealities, and systematic offsets not captured in the numerical model. Future work should therefore incorporate synthetic noise or hybrid train-on-simulation--fine-tune-on-experiment protocols to bridge this gap.

\section{Methods}
Proposed method uses conductance maps as high-dimensional visual inputs for the ViT model. The conductance $G$ is calculated using the $S$-matrix formalism in the wide-band limit~\cite{Donarini2024} via the Weidenm\"uller formula~\cite{Bordin2025,christiansen2009} for $N=3$ QDs chain:
\begin{equation}
    S(E_F)=
    \begin{pmatrix}
    s^{ee} & s^{eh}\\
    s^{he} & s^{hh}
    \end{pmatrix}
    =\mathbf{1}-iW^\dagger(E_F-H+\frac{i}{2}WW^\dagger)^{-1}W,
\end{equation}
with the tunneling matrix defined as $W=\mathrm{diag}(1,0,1)\otimes\mathrm{diag}(\sqrt{\Gamma},\sqrt{\Gamma},-\sqrt{\Gamma},-\sqrt{\Gamma})$ and the dot-lead coupling $\Gamma=0.1$~meV. 
If we reshape the $S$-matrix accordingly $S=S_{n,p,s,n'\!,p'\!,s'}$ ($n=\mathrm{L,C,R}$ indexing dots, $p=1,2$ particle, and $s$ -- spin subspaces) then the respective reflection matrices: $s^{ee}_{s,s'}\!(i,j)=S_{i,1,s,j,1,s'}$ and $s^{he}_{s,s'}\!(i,j)=S_{i,2,s,j,1,s'}$ give differential conductance as, $ss^\dagger\equiv|s|^2$,
\begin{equation}
G_{ij}(E_F)\equiv\frac{\mathrm{d}I_i}{\mathrm{d} V_j}=2\delta_{ij}-\mathrm{tr}\!\left(|s^{ee}(i,j)|^2\right)+\mathrm{tr}\!\left(|s^{he}(i,j)|^2\right)
\end{equation}
in unit of $e^2\!/h$, $i,j=\mathrm{L,R}$ denoting left (L) or right (R) lead, and $E_F$ being the Fermi energy in the leads.
The input maps include 4 conductance $G_{ij}$ components: $G_\mathrm{LL}$, $G_\mathrm{LR}$, $G_\mathrm{RL}$, and $G_\mathrm{RR}$, where for instance $G_\mathrm{LL}=\frac{dI_\mathrm{L}}{dV_\mathrm{L}}$, with $I_\mathrm{L}$ denoting current through the left lead, and $V_\mathrm{L}$ is the bias voltage of the left lead. Similarly, other components can be defined by using different combinations of left (L) and right (R) leads. 
Noteworthy, each conductance map is defined by 2D plot of $G_{ij}$ as a function of some parameter and $E_F$.
We utilize 4 maps per each component: 3 for \(\mu_{i=\mathrm{L},\mathrm{C},\mathrm{R}}\) variation and 1 for \(V_\mathrm{Z}\) variation: 16 maps in total serving as the input. Fig.~\ref{fig:default_conductance} shows the conductance maps ($G_{\mathrm{LL}}$ component) for the reference parameters $P_0$, highlighting the emerged zero-biased peaks.

The $\mathcal{M}$ metric, utilized to train the PINNAT model, is the following:
\begin{align}\label{eq:majoranization}
\mathcal{M}(H(P))&=
\frac{p_0}{2}\max\!\left[0,\,2\,m_0-\sum_{i>1} m_i\right],\\
m_i&=|\langle\psi_i|M\rangle|e^{-|E_i|/\epsilon},\nonumber\\
p_i&=2\max\!\left[0,\,4|u_i|^2|v_i|^2 - \tfrac12\right],\nonumber
\end{align}
with $\{\psi_i,E_i\}$ being the collection of $H(P)$ eigenpairs. The index
$i$ enumerates the subsequent eigenvalues: both $m_i$ and $p_i$ are sorted in ascending order by $|E_i|$. 
Eigenstates projection $m_i$ on the left (or equivalently right) Majorana mode \(M = \gamma_\mathrm{L}\) is weighted by their distance from zero energy $|E_i|$ with some threshold \(\epsilon=0.1\)~meV. Second factor -- $p_i$ quantifies electron-hole symmetry: \(u_i,v_i\) are the electron-hole components of eigenvectors $\psi_i$. By the $H$ symmetry $m_0=m_1$, and also $p_0=p_1$. For ideal MZM pair we should have $m_0=m_1=1$ and $p_0=p_1=1$, giving $\mathcal{M}=1$. 
The form of the metric, Eq.~\ref{eq:majoranization}, is designed to discriminate trivial zero modes such as ABS, which discrimination from MZMs is described e.g. in Ref.~\cite{Bordin2025, Liu2017, Prada2020}. Especially, the $-\sum_{i>1}m_i$ term penalizes the presence of trivial zero modes, even when the topological ones are there as well. 

Although the metric defined in Eq.~(5) contains piecewise operations such as $\max(0,\cdot)$ and sorting by $|E_i|$, gradients required for training are computed using automatic differentiation provided by the \emph{autograd} framework. Such piecewise functions are handled using standard subgradient rules (analogous to ReLU activations), so that they propagate whenever the argument of the $\max(0,\cdot)$ operation is positive. The eigenvalues and eigenvectors are obtained from a differentiable eigensolver applied to the Hamiltonian. Degeneracies and level crossings occur only on sets of measure zero in parameter space and did not cause practical issues during training. However, eigenvector-based gradients may become numerically ill-conditioned near degeneracies or for larger matrices, which can introduce noise in the optimization signal. This potential limitation is discussed further in Appendix~C.

For the visualization of the spectra in Figs.~5 and~7, two auxiliary quantities are computed from the normalized BdG eigenvectors $\psi_i$, utilizing their electron and hole components $u_i$, $v_i$. The site occupation of dot $n$ for eigenstate $i$ is defined as
\begin{equation}
O^{(i)}_{n} = |u_{i,n}|^2 + |v_{i, n}|^2.
\end{equation}
The quantity referred to as edge occupations is the total weight on the outer dots,
\begin{equation}
O_{edge}^{(i)} = O^{(i)}_\mathrm{L} + O^{(i)}_\mathrm{R}.
\end{equation}
In addition, electron-hole symmetry is defined as the difference between the total electron and hole weights
\begin{equation}
S_{\mathrm{eh}}^{(i)} = \sum_{n}\left(|u_{i,n}|^2 - |v_{i,n}|^2\right).
\end{equation}

In our work we adapt typical ViT architecture~\cite{dosovitskiy2021} adjusted to process 16-channel $50\times 50$ input tensor (representing all generated conductance map components) and returning vector of the Hamiltonian parameter corrections $\delta P$. The architecture hyperparameters such as number of attention heads (4), number of attention blocks (6), hidden size (256) and patch size (1) are tuned to decrease value of $\mathcal{L}$ after 100 epochs below given threshold $\tau=-0.7$.

During single epoch of training we iterate through $10\,000$ independent $H$ samples generated using Eq.~\ref{hamiltonian}, varying parameters \(\{\mu_n, t_n, V_\mathrm{Z}, \Delta_i\}\) in range \([0, 1]\) \(\mathrm{meV}\) and \(\lambda_n\) in range \([0, \pi]\). The training is proceeded until convergence of $\mathcal{L}$, Eq.~\ref{eq:loss}, is observed -- usually between 150 and 200 epochs.

\section{Conclusions}

In this work, we demonstrated that ViT-based neural network physics-informed by $\mathcal{M}$ metric, trained on conductance maps can effectively autotune QDs chain Hamiltonian toward the emergence of MZMs. Our results show that the proposed framework succeeds in correcting both global deviations of system parameters -- those that shift all QDs uniformly -- and local parameter noise that affects individual QDs independently. In both regimes, PINNAT consistently increases the Majorana metric $\mathcal{M}$, thereby restoring the formation of zero-energy edge states where possible.

Proposed framework merges quantum transport simulation with ML-based parameter feedback for topological phase tuning. By integrating experimental observables (conductance maps) as model input and theoretical descriptors ($\mathcal{M}$ measure) as physics-informed targets for model training, our approach paves the way for autonomous tuning into robust topological regimes in noisy mesoscopic systems.

\section*{Acknowledgements}
We would like to thank M. Marga\'nska-\L{}y\.zniak for interesting discussions and helpful comments on the manuscript.
We gratefully acknowledge Polish high-performance computing infrastructure PLGrid (HPC Centers: ACK Cyfronet AGH) for providing computer facilities and support within computational grant no. PLG/2025/018433.

\appendix
\section{Appendix A: Analytical considerations for 3 QDs model}

Let's start with the Rashba-Zeeman-BCS Hamiltonian, Eq.~1, written in so-called Nambu spinor representation $\hat{\Psi}_n=(c_{n\uparrow},c_{n\downarrow},c^\dagger_{n\uparrow},c^\dagger_{n\downarrow})^T$: $H=\frac{1}{2}\sum_{nm}\hat{\Psi}^\dagger_n H^\mathrm{BdG}_{nm}\hat{\Psi}_m$, where $H^\mathrm{BdG}$ is a first-quantized Hamiltonian, also called the Bogoliubov-de Gennes (BdG) Hamiltonian. It's explicit matrix form reads:
\begin{align}\label{eq:hbdg}
&H^\mathrm{BdG}=\nonumber\\
&\sum_{n=1}^{N}
\begin{pmatrix}
-\mu_n+V_\mathrm{Z} & 0 & 0 & \Delta_n e^{i\phi_n}\\
0 & -\mu_n-V_\mathrm{Z} & -\Delta_n e^{i\phi_n} & 0\\
0 & -\Delta_n e^{-i\phi_n} & \mu_n-V_\mathrm{Z} & 0\\
\Delta_n e^{-i\phi_n} & 0 & 0 & \mu_n+V_\mathrm{Z}
\end{pmatrix}_{\!n,n}\nonumber\\
+&\sum_{n=1}^{N-1}
t_n\begin{pmatrix}
e^{i\boldsymbol{\lambda}_n \cdot \boldsymbol{\sigma}} & 0\\
0 & -e^{-i\boldsymbol{\lambda}_n \cdot \boldsymbol{\sigma}^T}
\end{pmatrix}_{\!n,n+1}+\mathrm{h.c.},
\end{align}
with the onsite (dot) block $(\quad)_{n,n}$ and the offsite hopping (inter-dot) block $(\quad)_{n,n+1}$---both represented by $4\times 4$ matrices. For a system consisting of three quantum dots ($N=3$), the full BdG Hamiltonian $H^\mathrm{BdG}$ is a $12\times 12$ matrix.

Our goal is to simulate the Kitaev chain (KC) model, i.e., to tune the $H^\mathrm{BdG}$ (parameters) to mimic KC Hamiltonian form.
The first step is that each
QD should have a pair of zero energy levels $E_n=0$. Diagonalizing a (separated) single dot block $(\quad)_{n,n}$ yields the condition for this to happen: $V_\mathrm{Z}=\sqrt{\Delta^2_n+\mu^2_n}$ or equivalently $\Delta_n=\sqrt{V^2_\mathrm{Z}-\mu^2_n}$.
When this condition is satisfied, each dot has a pair of fermionic excitations $a_n, a^\dagger_n$, where
\begin{equation}
a_n=\frac{1}{\sqrt{2V_\mathrm{Z}}}\left(e^{-\frac{i\phi_n}{2}}\sqrt{V_\mathrm{Z}+\mu_n}\,c^\dagger_{n\uparrow}-e^{\frac{i\phi_n}{2}}\sqrt{V_\mathrm{Z}-\mu_n}\,c_{n\downarrow}\right)\!.
\end{equation}
The energy of $a_n$ ($a^\dagger_n$) is zero. Two other modes are separated by the energy $\pm2V_\mathrm{Z}$.
Let's now assume that the hopping is much smaller than the
energy of the excited state, $t\ll2V_\mathrm{Z}$, we may project the Hamiltonian~(\ref{eq:hbdg}) onto the Hilbert space spanned by a set $\{a_1,a^\dagger_1,a_2,a^\dagger_2,\dots,a_N,a^\dagger_N\}$. To simplify the system definition, we take $\boldsymbol{\lambda}_n \cdot \boldsymbol{\sigma}=\lambda_n\sigma_y$, which follows from choosing the Rashba vector direction $\rho=\xi=\frac{\pi}{2}$.
The resulting projected Hamiltonian is:
\begin{equation}\label{eq:bdg_6}
\tilde{H}^\mathrm{BdG}=
\sum_{n=1}^{N-1}
\frac{t_n}{V_\mathrm{Z}}\begin{pmatrix}
T_n & -D_n\\
D_n & -T^\ast_n
\end{pmatrix}_{\!n,n+1}+\mathrm{h.c.},
\end{equation}
where $T_n=-(\mu_n\cos{\frac{\phi_{n+1}-\phi_n}{2}}-iV_\mathrm{Z}\sin{\frac{\phi_{n+1}-\phi_n}{2}})\cos{\lambda_n}$, and $D_n=-\Delta_n\cos{\frac{\phi_{n+1}-\phi_n}{2}}\sin{\lambda_n}$.
The goal is to simulate the KC model, which, e.g. for $N=3$ takes the form (in the same Nambu representation $(a_1,a^\dagger_1,a_2,a^\dagger_2,a_3,a^\dagger_3)$ as $\tilde{H}^\mathrm{BdG}$):
\begin{equation}\label{eq:kch_6}
H^\mathrm{KC}=
\begin{pmatrix}
\mu_1 & 0 & t'_1 & -\Delta'_1 & 0 & 0\\
0 & -\mu_1 & \Delta'_1 & -t'_1 & 0 & 0\\
t'_1 & \Delta'_1 & \mu_2 & 0 & t'_2 &-\Delta'_2e^{i\Phi}\\
-\Delta'_1 & -t'_1 & 0 & -\mu_2 & \Delta'_2e^{i\Phi} & -t'_2\\
0 & 0 & t'_2 & \Delta'_2e^{-i\Phi} & \mu_3 & 0\\
0 & 0 & -\Delta'_2e^{-i\Phi} & -t'_2 & 0 & -\mu_3
\end{pmatrix}\!.
\end{equation}
KC Hamitonian, Eq.~\ref{eq:kch_6}, at the \emph{sweet-spot} $\mu_n=0$, $t'_n=\Delta'_n$ possesses two MZMs.
For the Rashba–Zeeman
$s$-wave Hamiltonian $\tilde{H}^\mathrm{BdG}$, Eq.~\ref{eq:bdg_6}, to simulate $H^\mathrm{KC}$, Eq.~\ref{eq:kch_6}, at the sweet-spot, the following conditions must be satisfied (assuming $\Phi=0$):
\begin{align}
\phi_{n+1}&=\phi_n,\nonumber\\
\mu_n\cos{\lambda_n}&=\Delta_n\sin{\lambda_n}.
\end{align}
The latter yields $\lambda_n=\arctan(\frac{\mu_n}{\Delta_n})$ which, together with
$\mu_n=\sqrt{V^2_\mathrm{Z}-\Delta^2_n}$, gives the approximate position of the parameters sweet-spot for the QDs Hamiltonian $H$.
Both conditions are also discussed in the adaptive tuning proposal of Fulga \textit{et al.}~\cite{Fulga2013}, as well as in the recent experimental realization of short KCs of Bordin \emph{et al.}~\cite{Bordin2026}. 

\section{Appendix~B: Analyzing tuning procedure for hopping $t$ parameter}

\begin{figure*}[!htb]
    \includegraphics[width=.5\linewidth]{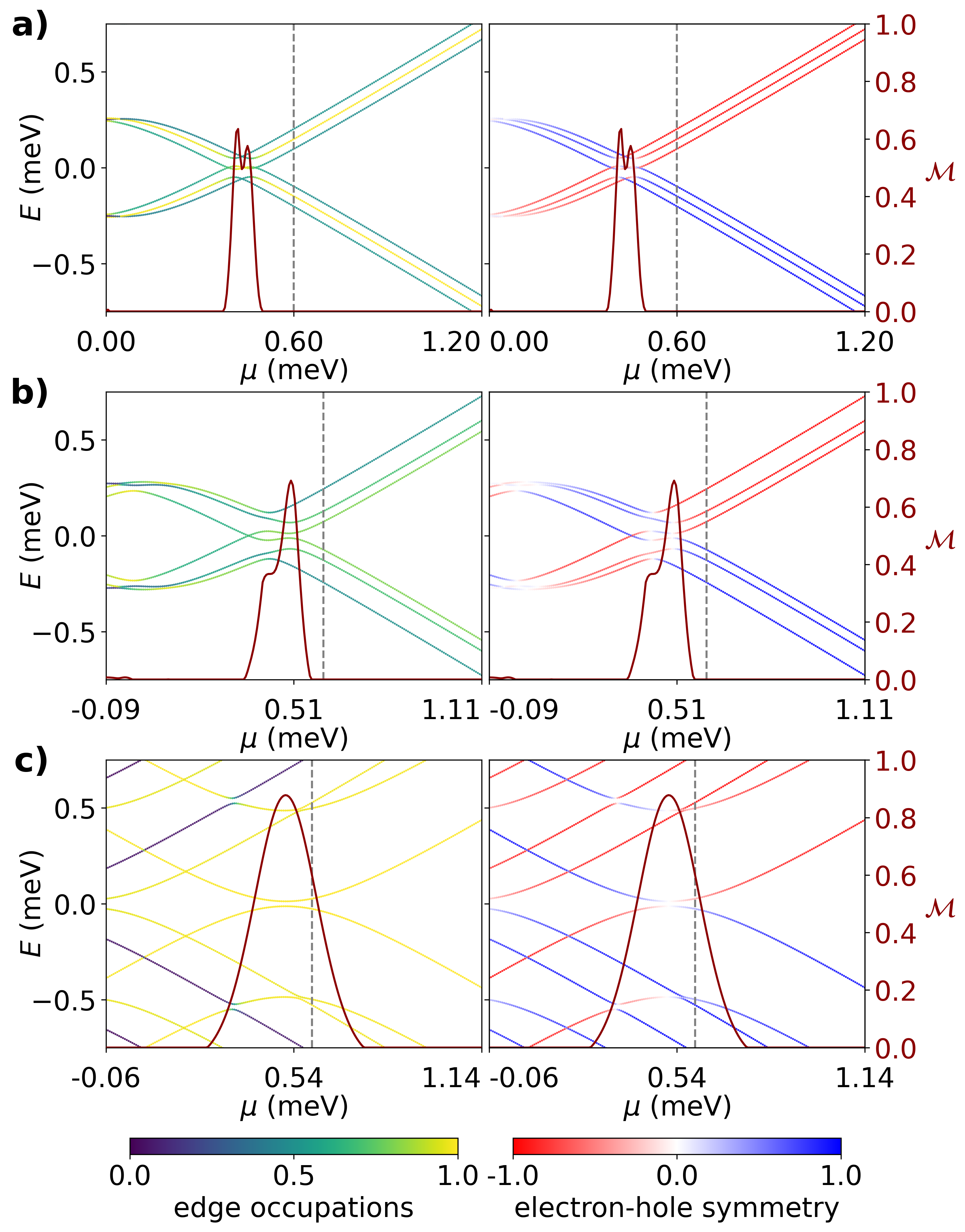}
    \caption{Three-QD chain Hamiltonian $H$ as a function of the offset $\mu$ added to local potentials: $\mu_n\rightarrow\mu_n+\mu-0.6$~meV (vertical dashed line marks the reference $\mu=0.6\,\mathrm{meV}$). Additionally (twin $y$-axis) $\mathcal{M}$ metric is plotted, and eigenvalues are colored (left column) with edge occupations $O_{edge}^{(i)}$, or (right column) electron-hole symmetry $S_{\mathrm{eh}}^{(i)}$ defined in the Methods section. Plots are presented for (a) reference parameters modified by setting $t_0=0.06\,\mathrm{meV}$ and $t_1=0.06\,\mathrm{meV}$; (b) parameters from (a) with PINNATuned \(\{\mu_n, t_n, \lambda_n\}\); (c) parameters from (a) with PINNATuned \(\{\mu_n, V_\mathrm{Z}\}\).}
    \label{fig:sample_t}
\end{figure*}

Here we present an additional example of an autotuning procedure for the parameter $t$. In Fig.~\ref{fig:sample_t}(a), we plot the $H$ eigenspectrum for detuned parameters $t_0 = t_1 = 0.06$~meV, fixed to reduce $\mathcal{M}$ value at reference $\mu=0.6$~meV. It is visible that such a setting of $t_0$ and $t_1$ does not make zero modes vanish completely, but instead moves them to a lower $\mu$ regime. Simultaneously, the zero-energy states are far less gapped from the rest of the spectrum than for the reference $t=0.25$~meV setup.

In the next step we try to tune the parameters $\{\mu_n,t_n,\lambda_n\}$ using the NN model, as presented in Fig~\ref{fig:sample_t}(b). We observe that PINNAT adjusts $\mu_n$ reducing the average $\langle\mu_n\rangle = 0.51$, thereby shifting the energy offset and restoring the previously observed zero modes. Additionally, further setting $t_0=0.11$~meV, $t_1=0.1$~meV, $\lambda_0=0.38\,\pi$ and $\lambda_1=0.26\,\pi$ parameters allows NN to increase the $\mathcal{M}$ value and as a result, increase the topological gap -- separation between zero modes and the bulk spectrum.

Finally, we can adjust the detuning in $t_0$ and $t_1$ using the second version of PINNAT model that sets $\{\mu_n,V_\mathrm{Z}\}$ parameters -- Fig.~\ref{fig:sample_t}(c). Similarly like before, $\mu_n$ are shifted so that their average value is decreased: $\langle\mu_n\rangle = 0.54$~meV, and zero modes emergence is observed. Moreover, the NN model proposes to set $V_\mathrm{Z} = 0.24$~meV, yielding a higher value of $\mathcal{M}$ and an even larger topological gap than that in Fig.~\ref{fig:sample_t}(b), where the model tunes $t_n$ and $\lambda_n$.

Nevertheless, similarly to the example presented in the main text, we can observe that both models are capable of correcting given parameters to produce $H$ with emergent zero-energy states, which are likely to maintain MZMs properties.

\subsection{Appendix~C: Scaling the system size}

\begin{figure}[b]
    \includegraphics[width=.99\linewidth]{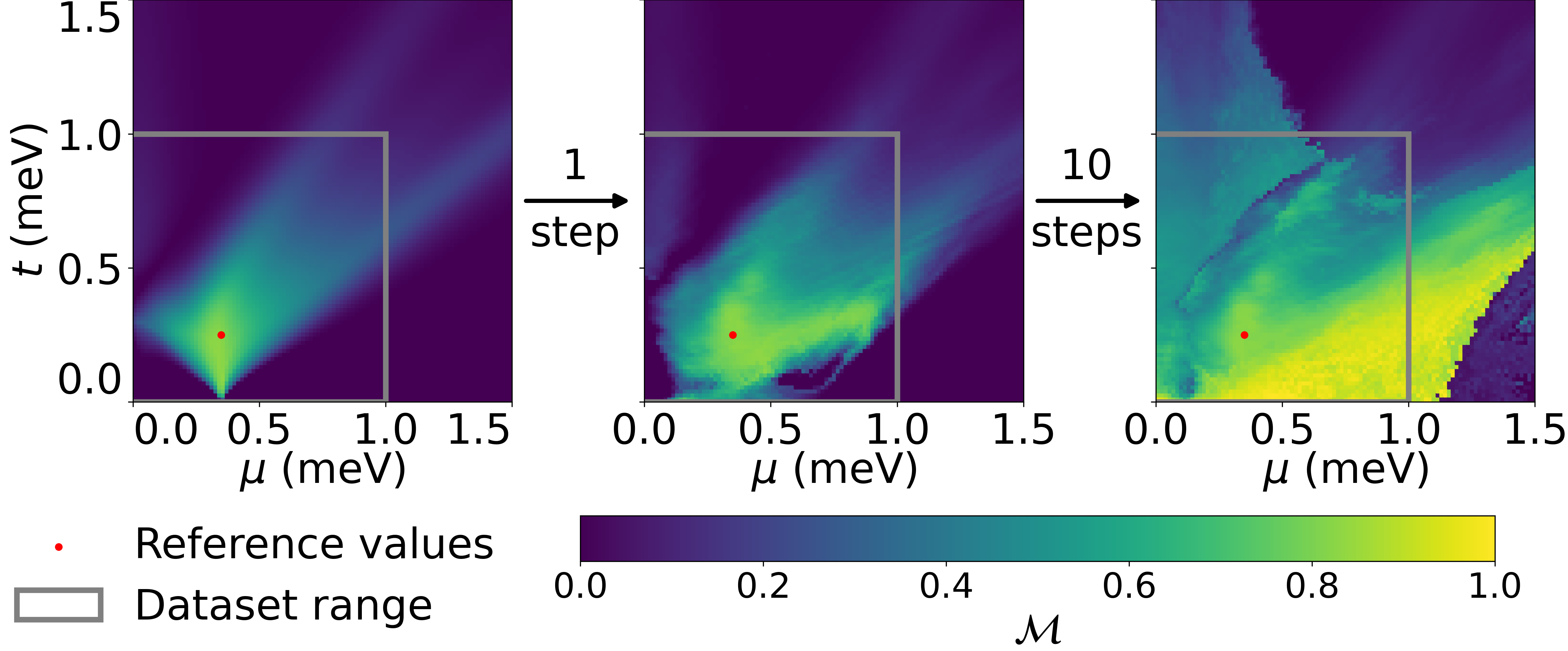}
    \caption{Iterative autotuning procedure for model adjusting \(\{\mu_n, V_\mathrm{Z}\}\) for system composed of 7 QDs. In the left plot, we present $\mathcal{M}$ map before tuning. Then, in the center, parameters are tuned with a single step of NN corrections. Finally, in the right plot, there is a map plotted for parameters tuned within 10 subsequent steps of NN corrections.}
    \label{fig:m_map_autotuning_7_dots}
\end{figure}

While analyzing results for a quantum chain composed of 3 QDs only proves the conceptual validity of the proposed approach, the natural question arises whether it is possible to repeat a similar procedure for larger systems. For this reason, we manually tune the Rashba-Zeeman-BCS Hamiltonian, Eq.~1, to realize MZMs for $N=7$ quantum dots. Such reference parameters include $\mu=0.35\,\mathrm{meV}$, $t=0.25\,\mathrm{meV}$, $\lambda=0.27\,\pi$, $\rho=\xi=\frac{\pi}{2}$, $V_\mathrm{Z}=0.5\,\mathrm{meV}$, and $\Delta=0.35\,\mathrm{meV}$.
In contrast to 3-QDs systems, to train the PINNAT model, for the 7-QDs chain, we could not use entirely random parameter sampling due to a much larger parameter space and far less number of samples with $\mathcal{M} > 0$. Instead, we generated 10\,000 samples, randomly shifting the parameters from the reference values with the restriction that $\mathcal{M}$ value is larger than $\tau=0.7$. Subsequently, in each training epoch, the samples are independently randomized to decrease the value of $\mathcal{M}$. For half of the samples, $\mathcal{M}$ is reduced to 0, and for the remaining 5\,000 samples, $\mathcal{M}$ is decreased to the interval $(0,\tau]$. Next, a set of 16 conductance maps is determined: to reduce the conductance tensor size we use only the $G_\mathrm{LL}$ and $G_\mathrm{RR}$ pair as a function of 8 parameters: \(\mu_{n}\), for $n$ iterating over each dot, and \(V_\mathrm{Z}\) (16 maps in total). Finally, the PINNAT model is trained to increase the value of $\mathcal{M}$ by adjusting the \(\{\mu_n, V_\mathrm{Z}\}\) parameters.  

The results of training the model are shown in Fig.~\ref{fig:m_map_autotuning_7_dots}, where the $\mathcal{M}$ metric is plotted for the previously introduced 10-step autotuning procedure. In contrast to the 3-QDs system, Fig.~4(b), the PINNAT model trained to produce corrections for 7-QDs chain tends to tune the system to a lesser extent. However, the increase in $\mathcal{M}$ value is still visible even after a single step of corrections. Moreover, within the 10 steps, the PINNAT model demonstrates its ability to correct parameter detuning in the whole training distribution and also shows the potential of generalization beyond the training regime.

While these results prove that it is possible to scale the PINNAT model with an autotuning approach to larger systems, several limitations to the scaling should be highlighted.
In particular, the dataset construction requires filtering Hamiltonians that exhibit emerging Majorana modes, while the training procedure involves repeated direct diagonalization within the loss evaluation (via the $\mathcal{M}$ metric) and the computation of conductance maps. As a result, both the data generation time and the training time scale approximately exponentially with the number of quantum dots in the system. This behavior is illustrated in Table~\ref{tab:scaling}, which summarizes the observed computational costs for systems of different sizes. While small systems can be handled almost instantaneously, the computational cost increases rapidly. For example, generating the dataset for a 10-dot system requires $\sim17$ hours, and a single training epoch takes $\sim5.5$ hours. Therefore, extending the method to larger systems will likely require additional optimizations, such as limiting the number of input conductance maps, reducing the number of tuned parameters, narrowing the parameter search ranges, or constructing a more time-effective target function.

\begin{table}[b]
\centering
\begin{tabular}{c c c}
\hline
\makecell{\textbf{Number}\\\textbf{of dots}} &
\makecell{\textbf{Data generation}\\\textbf{time [h]}} &
\makecell{\textbf{Training time}\\\textbf{per epoch [h]}} \\
\hline
3  & online & 0.28 \\
5  & 2  & 0.42 \\
7  & 5   & 1.17 \\
10 & 17  & 5.50 \\
\hline
\end{tabular}
\caption{Observed scaling of data generation and model training times with system size.}
\label{tab:scaling}
\end{table}

Additionally, although the eigendecomposition used in the loss evaluation is differentiable almost everywhere, the gradients associated with eigenvectors can become numerically ill-conditioned near degeneracies or level crossings. This effect may introduce noise in the backpropagated gradients and can potentially limit the scalability of eigenvalue-based learning procedures. Various approaches have been proposed to mitigate this issue in differentiable eigensolvers~\cite{Wang2019,Tianbo2024}. However, in our experiments, we observed stable and well-behaved training for systems up to 10 quantum dots (corresponding to $40\times40$ Hamiltonian matrices), with no noticeable degradation of the optimization process.

\subsection{Appendix~D: Comparison of tuned parameters with Majorana sweet-spot conditions}

\begin{table*}[t]
\centering
\caption{Comparison of reference and PINNAT parameter sets for the
three-dot chain. Energies are given in meV. In the presented sets $\Delta_n$ does not change with $n$: $\Delta_n=\Delta$. 
}
\label{tab:param_comparison}

\begin{tabular}{lcccccccccccccc}
\toprule
Configuration
& $\langle\mu_n\rangle$
& $V_\mathrm{Z}$
& $\Delta$
& $t_1$
& $t_2$
& $\lambda_1$
& $\lambda_2$
& $\frac{V_\mathrm{Z}}{\max_n{t_n}}$
& $\frac{V_\mathrm{Z}}{\sqrt{\Delta^2+\langle\mu_n\rangle^2}}$ 
& $\frac{\langle\lambda_n\rangle}{\arctan\left(\frac{\langle\mu_n\rangle}{\Delta}\right)}$\\
\midrule

Reference parameters
& 0.60
& 0.50
& 0.25
& 0.25 & 0.25
& $0.27\pi$ & $0.27\pi$
& 2.0 & 
0.77 & 
0.72\\

PINNAT $\{\mu_n,t_n,\lambda_n\}$ (Fig.~5c)
& 0.54
& 0.50
& 0.25
& 0.20 & 0.25
& $0.67\pi$ & $0.41\pi$
& 2.0 & 
0.83 & 
1.49\\

PINNAT $\{\mu_n,V_\mathrm{Z}\}$ (Fig.~5d)
& 0.58
& 0.31
& 0.25
& 0.25 & 0.25
& $0.62\pi$ & $0.41\pi$
& 1.24 & 
0.49 & 
1.39\\

PINNAT $\{\mu_n,t_n,\lambda_n\}$ (Fig.~7b)
& 0.51
& 0.50
& 0.25
& 0.11 & 0.10
& $0.38\pi$ & $0.26\pi$
& 4.55 & 
0.88 & 
0.90\\

PINNAT $\{\mu_n,V_\mathrm{Z}\}$ (Fig.~7c)
& 0.54
& 0.24
& 0.25
& 0.06 & 0.06
& $0.27\pi$ & $0.27\pi$
& 4.0 & 
0.4 & 
0.75 \\
\bottomrule
\end{tabular}
\end{table*}

To facilitate comparison with previously proposed parameter regimes for Majorana zero modes in few-site Kitaev chains, Table~\ref{tab:param_comparison} lists representative sets of PINNAT parameters together with the reference values used in the simulations. In addition to the raw parameters, dimensionless ratios are reported that quantify proximity to the sweet-spot relations discussed in Appendix A.

The PINNAT configurations consistently fall within the expected Majorana-supporting regime. In particular, the Zeeman energy exceeds the inter-dot couplings in all cases, as indicated by $V_\mathrm{Z} / \max_n t_n > 1$, often by a significant margin. This places the system in a regime where Zeeman splitting dominates over inter-dot hybridization, which is favorable for the emergence of topological phases.

Furthermore, the ratio $V_\mathrm{Z} / \sqrt{\Delta^2 + \langle \mu_n \rangle^2}$ remains of order unity, indicating that the condition $V_\mathrm{Z} = \sqrt{\Delta^2 + \mu_n^2}$ for having a single fermionic pair at zero energy per dot is approximately satisfied.
Finally, the relation between the Rashba spin-orbit length and the onsite potential, captured by $\langle \lambda_n \rangle /\arctan\frac{\langle \mu_n \rangle}{\Delta}$, also remains of order unity across the considered configurations, indicating consistency with the analytical condition derived in Appendix~A.

Overall, while the analytical sweet-spot conditions are not exactly fulfilled, the PINNAT parameters systematically lie in their approximate regime of validity, demonstrating that the autotuning procedure identifies physically consistent Majorana-supporting configurations.

\bibliography{main}

\end{document}